\documentclass[%
preprint,
 amsmath,amssymb,
 aps,
]{revtex4-2}

\usepackage{graphicx}

\begin{document}

\title{Particle collisionality in scaled kinetic plasma simulations}

\author{S. R. Totorica}
\email{totorica@princeton.edu}
\affiliation{
 Department of Astrophysical Sciences, Princeton University, Princeton, NJ 08544, USA
}
\affiliation{Department of Astro-fusion Plasma Physics (AFP), Headquarters for Co-Creation Strategy, National Institutes of Natural Sciences, Tokyo 105-0001, Japan}

\author{K. V. Lezhnin}
\affiliation{
 Princeton Plasma Physics Laboratory, Princeton, NJ 08540, USA
}%

\author{W. Fox}
\affiliation{
 Department of Astrophysical Sciences, Princeton University, Princeton, NJ 08544, USA
}%
\affiliation{
 Princeton Plasma Physics Laboratory, Princeton, NJ 08540, USA
}%

\begin{abstract}
Kinetic plasma processes, such as magnetic reconnection, collisionless 
shocks, and turbulence, are fundamental to the dynamics of astrophysical 
and laboratory plasmas. Simulating these processes often requires
particle-in-cell (PIC) methods, but the computational cost of fully kinetic 
simulations can necessitate the use of artificial parameters, such as a 
reduced speed of light and ion-to-electron mass ratio, to decrease expense.
While these approximations can preserve overall dynamics under specific 
conditions, they introduce nontrivial impacts on particle collisionality
that are not yet well understood. In this work, we develop a method to 
scale particle collisionality in simulations employing such approximations.
By introducing species-dependent scaling factors, we independently adjust 
inter- and intra-species collision rates to better replicate the 
collisional properties of the physical system. Our approach maintains the 
fidelity of electron and ion transport properties while preserving critical 
relaxation rates, such as energy exchange timescales, within the limits of
weakly collisional plasma theory. We demonstrate the accuracy of this 
scaling method through benchmarking tests against theoretical relaxation 
rates and connecting to fluid theory, highlighting its ability to retain 
key transport properties. Existing collisional PIC implementations can be 
easily modified to include this scaling, which will enable deeper insights
into the behavior of marginally collisional plasmas across various contexts.

\end{abstract}

\maketitle

\section{Introduction}

Kinetic plasma processes such as magnetic reconnection, collisionless
shocks, and turbulence play key roles in the dynamics of many astrophysical 
and laboratory plasmas \cite{Blandford2014}.  The microscale, nonequilibirum dynamics associated
with these processes can produce non-Maxwellian particle distributions
such as power-law tails of nonthermal energetic particles, as well as
significantly influence the large-scale fluid motions of the system
through their complex coupling to global scales. For the plasma
conditions of many systems of interest in space physics and astrophysics,
the timescales for collisional processes are often sufficiently long
compared to those of collective plasma processes that they may be neglected
entirely, and the system can be accurately modeled in the collisionless
plasma approximation. However, for most laboratory plasmas such as those 
produced in tokamaks or high-energy-density (HED) facilities, as well as 
certain natural environments including accretion disks, collisional 
processes often have important impacts on the dynamics of the system and the
details of the resulting particle distributions.

Investigations of kinetic plasma dynamics typically require large-scale 
numerical simulations, with one of the most powerful and widely adopted
approaches being the first-principles particle-in-cell (PIC) method \cite{Birdsall}.
The interparticle force calculation of the PIC algorithm involves 
interpolation between continuous simulation particle positions and discrete
field quantities defined on a grid that covers the spatial domain.
The grid effectively gives the particles a finite shape, which reduces the 
collisionality of the system to negligible levels at sufficient particle
resolution. Finite collisional effects can be incorporated into PIC
simulations by extending the base algorithm with a binary Monte-Carlo 
Coulomb collision operator, which periodically scatters nearby simulation 
particles with momentum-dependent deflections that reproduce theoretical 
relaxation rates \cite{Takizuka1977}. This method has been successfully applied to gain insight 
into a variety of weakly collisional plasma systems from astrophysics to the
laboratory.

The accuracy of PIC simulations comes with a large computational
expense that limits the spatial and temporal scales that
can be modeled, and often makes it necessary to reduce the simulated
system to two or even one spatial dimension. For studies of 
electromagnetic plasma dynamics that capture both the ion and electron
kinetic scales, such as magnetic reconnection in planetary magnetospheres
or HED laboratory astrophysics experiments, fully kinetic
simulations in multiple dimensions are typically only possible when further 
approximations are made through the use of artificial physical 
parameters. In particular, increasing the electron mass (reducing the ion
to electron mass ratio) and reducing the speed of light in the simulation 
can drastically reduce the computational expense of modeling a given 
physical system. This comes with the trade-off of reducing the separation 
between electron and ion scales and/or electrostatic and electromagnetic 
scales. However, as long as adequate scale separation is maintained, the 
overall dynamics of the system can be expected to be preserved, and these 
approximations have been routinely used in PIC simulations for decades 
\cite{Pritchett2001,Hesse2001CollisionlessModeling,Fox2018KineticPlasmas}. For
many applications, in particular when capturing three-dimensional spatial
dynamics, these approximations will remain unavoidable for the foreseeable 
future.

The use of artificial physical parameters has nontrivial implications
for the collisionality of the simulated system that have not yet been fully 
investigated, in particular for the mass ratio. Past studies have 
compensated for the artificial physical parameters by scaling the collisionality
to preserve the properties of electron transport in the simulated system \cite{Daughton2009,Daughton2009a,Fox2011a,Fox2012a,Le2015}. 
However, in the typical approach of applying the same scaling to all types
of collisions, this will alter the ion transport properties compared to the
physical system. Similarly, if ion collisions are of primary interest,
matching the ion collisionality to the physical system will alter
that of the electrons. To improve on this, below we make a detailed 
examination of the effects of artificial physical parameters on the 
different types of inter- and intra-species collisions, and develop a new 
procedure to separately scale the different types of collisions to better 
retain the collisional properties of all plasma
species. We find that by introducing species-dependent scaling factors to the collision rates, a larger number of collisional processes can be properly reproduced. The new scaling applies to systems with nonrelativistic 
temperatures that satisfy $m_{i} T_{e} / m_{e} T_{i} \gg 1$, and although 
this study focuses on PIC simulations, it is equally valid for
other kinetic simulation methods such as Eulerian Vlasov codes. Existing 
collisional PIC implementations may be easily modified to incorporate the 
proposed scaling factors, which will improve the accuracy of modeling 
marginally collisional laboratory and astrophysical plasmas. 

\section{Scaling a kinetic simulation to reduce computational expense}

The scaling of a PIC simulation using artificial physical parameters 
to reduce computational expense can be parameterized using two 
scaling factors, $K_{c} = c_{sim} / c_{phys}$, describing the ratio 
of the artificially low speed of light in the simulated system to 
that of the physical system, and $K_{m} = m_{sim} / m_{phys}$, 
describing the ratio of the artificially heavy electron mass in the 
simulated system to that of the physical system. The electron and ion 
temperatures $T_{e}$ and $T_{i}$, bulk plasma flow
velocities $V_{fl}$, initial magnetic fields, electron and ion densities
$n_{e}$ and $n_{i}$, and characteristic length scale $L$ are assumed to be unchanged in the scaled system.
An alternative approach is to work in terms of increasing the 
simulation particle velocities compared to the speed of light and reducing 
the ion mass, however the two points of view are effectively equivalent, and
using the heavy electron mass and the slow speed of light is more intuitive
for understanding the impacts of the scaling on the collisionality of the
system. To preserve the ion scale properties of the system, the 
characteristic length scale in terms of the ion inertial length
($L / (c/\omega_{pi})$) and ion gyroradius ($L/\rho_{i}$) should be
matched between the scaled and physical systems. These ratios can be 
preserved in the scaled system by reducing the elementary charge $e$ as
$e_{sim} = e_{phys} K_{c}$ in CGS units, or increasing the permittivity of free space
$\epsilon_{0}$ as $(\epsilon_{0})_{sim} = (\epsilon_{0})_{phys} K_{c}^{-2}$ in SI
units. The dynamical results of the two approaches are equivalent, and in this work
we will proceed in CGS units.

The computational expense of a PIC simulation can be estimated 
approximately as $\sim N_{t} \times N_{G} \times N_{ppc}$,
where $N_{t}$ is the number of timesteps, $N_{G}$ is the number of
grid cells, and $N_{ppc}$ is the number of particles-per-cell.
To prevent artificial heating from the finite grid instability,
the maximum grid spacing $\Delta x$ is limited to approximately
$\Delta x \lesssim \lambda_{D}$, where $\lambda_{D} = \sqrt{T/4\pi n e^{2}}$ is the Debye
length. To ensure the stability of electromagnetic waves, the maximum
timestep is limited to $\Delta t < (\Delta x / c) / \sqrt{d}$ where
$d$ is the spatial dimensionality of the system. The number of grid
cells and timesteps required in the simulation can be estimated
approximately as $N_{G} \propto (L / \lambda_{D})^{d}$ and
$N_{t} \propto (L/V)/(\lambda_{D}/c)$. The ratio of the 
computational expense of simulating the system with scaled versus
physical parameters can then be estimated approximately as
$\frac{(\mathrm{Expense})_{sim}}{(\mathrm{Expense})_{phys}} \approx (K_{c})^{2+d}$ for $d$ spatial dimensions, demonstrating the 
possibility for drastic reductions in computational expense by 
reducing the speed of light in the scaled system.

Counterintuitively, there is no direct dependence of the computational 
expense on the electron mass, and therefore changing the electron 
mass alone does not directly lead to any computational savings. The 
importance of increasing the electron mass (reducing the ion to 
electron mass ratio) in the scaled system is that this allows further
reductions in the speed of light before electrons become artificially
relativistic, increasing the amount of computational savings that
can be attained. Assuming the physical electron temperature is 
nonrelativistic, retaining a nonrelativistic temperature in the
scaled system ($(T_{e} / m_{e} c^{2})_{sim} \ll 1$) limits the
possible choices of scaling parameters to those that satisfy
$K_{c}^{2} K_{m} \gg (T_{e} / m_{e} c^{2})_{phys}$. Increasing
$K_{m}$ allows the use of smaller values of $K_{c}$ while satisfying
this condition.

\section{Scaling collisionality to compensate for artificial speed of light and mass ratio}

 Test particle relaxation rates for particles of species $\alpha$ scattering off of species $\beta$ are given by \cite{Trubnikov1965,Huba2013NRLFORMULARY}
\begin{align}
  &\nu^{\alpha \backslash \beta}_{s} = \left ( 1 + m_{\alpha} / m_{\beta} \right ) \psi(x^{\alpha \backslash \beta})\nu^{\alpha \backslash \beta}_{0} \\
  &\nu^{\alpha \backslash \beta}_{\|} = \left [ \psi(x^{\alpha \backslash \beta}) / x^{\alpha \backslash \beta} \right ] \nu^{\alpha \backslash \beta}_{0} \\
  &\nu^{\alpha \backslash \beta}_{\perp} = 2 \left [ (1-1/2 x^{\alpha \backslash \beta}) \psi(x^{\alpha \backslash \beta}) +\psi^{\prime}(x^{\alpha \backslash \beta}) \right ] \nu^{\alpha \backslash \beta}_{0} \\
  &\nu^{\alpha \backslash \beta}_{\epsilon} = 2 \left [ \left ( m_{\alpha} / m_{\beta} \right )\psi(x^{\alpha \backslash \beta}) - \psi^{\prime}(x^{\alpha \backslash \beta}) \right ] \nu^{\alpha \backslash \beta}_{0} 
  \label{eq:relaxation_rates}
\end{align}
where the characteristic collision frequency is
\begin{equation}
    \nu^{\alpha \backslash \beta}_{0} = 4 \pi e_{\alpha}^{2} e_{\beta}^{2} \lambda_{\alpha \beta} n_{\beta} / m_{\alpha}^{2} v_{\alpha}^{3}
    \label{eq:nu0}
\end{equation}
and the preceding terms consisting of
\begin{align}
\psi(x) = \frac{2}{\sqrt{\pi}} \int_{0}^{x} dt \, t^{1 / 2} e^{-t} \quad\quad\quad&\quad\quad\quad \psi^{\prime}(x) = \frac{d \psi}{dx}
\end{align}
integrate over the particles of species $\beta$ that 
contribute to the scattering of a particle of species $\alpha$
with normalized energy $x^{\alpha \backslash \beta} = m_{\beta} v_{\alpha}^{2} / 2 k T_{\beta}$. Here $e_{\alpha}$ and $e_{\beta}$ are the respective charges of the particles of species $\alpha$ and $\beta$, $\lambda_{\alpha\beta}$ is the
Coulomb logarithm for collisions between particles of species $\alpha$ and $\beta$, $n_{\beta}$ is the density of particles of species $\beta$, and $m_{\alpha}$ and $v_{\alpha}$ are the mass and velocity of the incident particle of species $\alpha$. The relaxation rates $\nu_{s}$, $\nu_{\perp}$,
$\nu_{\|}$, and $\nu_{\epsilon}$ describe the slowing down,
transverse diffusion, parallel diffusion, and energy loss, respectively, for test particles with a given incident velocity.

In the Takizuka-Abe algorithm \cite{Takizuka1977} and its extensions which are commonly used in PIC codes, nearby
simulation particles are first randomly divided into pairs. Scattering
is then performed in the center-of-mass frame of each pair by randomly
sampling a scattering angle from a distribution that depends on the
relative velocity, reduced mass, and characteristic collision rate
(Eq. (\ref{eq:nu0})) for each pair. Different types of inter- and intera-species collisions can be separately scaled in PIC simulation
by introducing independent prefactors to the characteristic collision rates
used in the scattering angle calculation, as described below.

\subsection{Reduced speed of light}

As mentioned above, a commonly used approximation that can greatly 
reduce the expense of PIC simulations is using an artificially reduced 
speed of light, $K_{c} = c_{sim} / c_{phys}$, while keeping the
system size in terms of $L / d_{i}$ and $L / \rho_{i}$ constant by 
additionally reducing the elementary charge as $e_{sim} / e_{phys} = K_{c}$.
This has the effect of increasing the ratio of the Debye length to the
electron and ion skin depths as well as characteristic length scale of the system
\begin{equation}
    [\lambda_{D} /  (c/\omega_{pe})]_{sim} / [\lambda_{D} /  (c/\omega_{pe})]_{phys} = K_{c}^{-1}K_{m}^{-1/2}
    \label{eq:length_separation_e}
\end{equation}
\begin{equation}
    [\lambda_{D} / (c/\omega_{pi})]_{sim} / [\lambda_{D} /  (c/\omega_{pi})]_{phys} = K_{c}^{-1}
    \label{eq:length_separation_i}
\end{equation}
\begin{equation}
    [\lambda_{D} / L]_{sim} / [\lambda_{D} / L]_{phys} = K_{c}^{-1}
\end{equation}
which reduces the spatial resolution 
required to avoid numerical heating in PIC simulations.
Reducing the speed of light will also increase the ratio of the cyclotron
frequency to the plasma frequency for each species as
\begin{equation}
    [\omega_{ce} / \omega_{pe}]_{sim}/[\omega_{ce} / \omega_{pe}]_{phys} = K_{c}^{-1}K_{m}^{-1/2}
    \label{eq:time_separation_e}
\end{equation}
\begin{equation}
    [\omega_{ci} / \omega_{pi}]_{sim}/[\omega_{ci} / \omega_{pi}]_{phys} = K_{c}^{-1}
    \label{eq:time_separation_i}
\end{equation}
effectively artificially increasing the magnetization of
the simulation particles.

The particle relaxation rates given above are the result of nonrelativistic
collision theory and thus have no direct dependence on c.  However, the decrease
in the elementary charge $e$ that must occur to keep the system size in 
terms of $c/\omega_{pi}$ and $\rho_{i}$ constant with reduced speed of light
alters the characteristic collision frequency in the scaled system. 
To tune the collisionality of the scaled system, we
assume for simplicity that the Coulomb 
logarithms are equivalent to those of the physical system, and add the prefactor
$K_{\nu\alpha\beta}$ to the characteristic collision frequencies.
\begin{equation}
    (\nu_{0}^{\alpha \backslash \beta})_{sim} = (K_{\nu \alpha \beta})(\nu_{0}^{\alpha \backslash \beta})
\end{equation}
This allows for independently scaling each type of inter- and intra-species
collision without altering the collisionless dynamics.

The separation between electrostatic ($\lambda_{D}$, $\omega_{pe,i}$) and electromagnetic ($\rho_{e,i}$, $\omega_{ce,i}$, $c/\omega_{pe,i}$) scales is 
altered in the simulated system (Eqs. (\ref{eq:length_separation_e},\ref{eq:length_separation_i}) and (\ref{eq:time_separation_e},\ref{eq:time_separation_i})), so one must choose which of these scales to
use to set $K_{\nu \alpha \beta}$ to best mimic the physical system.  The
scaling of the collisionality in terms of these scales is given as follows.
\begin{equation}
(\nu^{\alpha \backslash \beta}_{0} / \omega_{c})_{sim} / (\nu^{\alpha \backslash \beta}_{0} / \omega_{c})_{phys} = K_{\nu \alpha \beta} K_{c}^{4}\end{equation}
\begin{equation}
(\nu^{\alpha \backslash \beta}_{0} / \omega_{p})_{sim} / (\nu^{\alpha \backslash \beta}_{0} / \omega_{p})_{phys} = K_{\nu \alpha \beta} K_{c}^{3}\end{equation}
As the scaling procedure described above was performed  to match dynamics at electromagnetic scales at the expense of altering the electrostatic dynamics, typically one is interested in
matching the collisional dynamics on the electromagnetic scales. In this case
one must choose $K_{\nu \alpha \beta} = K_{c}^{-4}$ (equivalent to simulating the collisional dynamics using the original, unscaled elementary charge $e_{phys}$).  
Collision frequencies will then be matched to the cyclotron frequency and 
mean free paths will be matched to the skin depth and gyroradius,
when compared to the physical system (assuming for now that the particle
masses are not altered from the physical system). For the electrostatic 
scales, collision frequencies in terms of the plasma frequency will be
increased by $K_{c}^{-1}$ and mean free paths in terms of the
Debye length will be reduced by a factor of $K_{c}$.  This may
impact electrostatic dynamics such as anomalous resistivity, and as will
be discussed below in Section \ref{sec:limitations} puts limits on the value of $K_{c}$ that can be used 
to model a given physical system. For systems
where electrostatic dynamics are of primary interest,
it is likely inappropriate to scale a simulation with
an artificial speed of light due to its alteration of 
the electrostatic dynamics. It is therefore
unclear if there are problems for which the scaling
$K_{\nu \alpha \beta} = K_{c}^{-3}$ would be preferred.

\subsection{Reduced mass ratio}

\subsubsection{Ion-ion and electron-electron collisions}

Another commonly used approximation for reducing the
computational expense of PIC simulations is using an 
artificially reduced ion to electron mass ratio.  In the
typical use case of matching the simulated and physical systems
at ion scales, this means the electron mass is effectively
increased as $K_{m} = (m_{e})_{sim}/(m_{e})_{phys}$.
Ion-ion collisions have no dependence on the electron mass,
so choosing $K_{\nu i i }= K_{c}^{-4}$ to account for the
reduced speed of light and elementary charge will still ensure that the 
ratios between all four ion-ion relaxation rates
(Eq. (\ref{eq:relaxation_rates})) and corresponding mean free paths 
have the same relation to the ion electromagnetic scales as in the physical 
system, as shown in the previous section.

For electron-electron collisions, we assume the particle velocities in 
the physical system are mapped to those of the simulated system
such that each corresponds to a Maxwellian distribution of temperature
$T_{e}$ for the electron mass in that system. Explicitly, a particle
with a velocity of $\kappa$ times the thermal velocity in the physical
system is mapped to the simulated system as
$\kappa \sqrt{T_{e}/(m_{e})_{phys}} \rightarrow \kappa\sqrt{T_{e}/(m_{e})_{sim}}$. This leaves the multiplicative functions
in front of the characteristic collision frequency 
$\nu_{0}^{e \backslash e}$ in the formulas for the four
relaxation rates invariant to changes in the electron
mass.  The properties of electron-electron collisions in the scaled
system are then altered in time and space through the changes in
$\nu_{0}^{e \backslash e}$ and $v_{t,e} = \sqrt{T_{e}/m_{e}}$.
The changes in various quantities of interest in the scaled system are:

\begin{equation}(\omega_{ce})_{sim}/(\omega_{ce})_{phys} = K_{m}^{-1}\end{equation}

\begin{equation}(\omega_{pe})_{sim}/(\omega_{pe})_{phys} = K_{c} K_{m}^{-1/2}\end{equation}

\begin{equation}(d_{e})_{sim}/(d_{e})_{phys} = K_{m}^{1/2}\end{equation}

\begin{equation}(v_{t,e})_{sim} / (v_{t,e})_{phys} = K_{m}^{-1/2}\end{equation}

\begin{equation}(\rho_{e})_{sim}/(\rho_{e})_{phys} = K_{m}^{1/2}\end{equation}

\begin{equation}(\nu_{0}^{e \backslash e})_{sim} / (\nu_{0}^{e \backslash e})_{phys} = K_{\nu ee} K_{c}^{4} K_{m}^{-1/2}\end{equation}
where $\rho_{e} = v_{t} / \omega_{ce}$ is the thermal electron gyroradius.
Choosing  $K_{\nu ee} = K_{c}^{-4} K_{m}^{-1/2}$ will
match the ratios of the electron-electron collision frequencies
to the electron cyclotron frequency, and mean free paths
to electron skin depth and gyroradius, to the ratios of the
physical system. While this matching is expected to preserve
important properties of electron trajectories and electron-scale
dynamics, it is important to note that the ratio of the electron mean
free path to the global system length scales will be altered as
$(\lambda_{mfp,e} / L)_{sim} / (\lambda_{mfp,e} / L)_{phys} = K_{m}^{1/2}$,
which may have implications for non-local transport effects 
\cite{Fox2018KineticPlasmas}.

\subsubsection{Electron-ion and ion-electron collisions}

For collisions between electrons and ions, the dependence of the
collision rates on the mass ratio is complex due to the functions
$\psi(x^{\alpha \backslash \beta})$ and $\psi^{\prime}(x^{\alpha \backslash \beta})$. However, for conditions where the
majority of the collisions satisfy $x^{\alpha \backslash \beta} \ll 1$ or
$x^{\alpha \backslash \beta} \gg 1$, one can consider the 
limiting forms of the relaxation rates which are amenable to scaling:

\begin{align}
  &\mathrm{For}\;x^{\alpha \backslash \beta} \ll 1 : \nonumber \\
&\nu^{\alpha \backslash \beta}_{s} \approx (4/3\sqrt{\pi}) \left ( 1 + m_{\alpha} / m_{\beta} \right ) (x^{\alpha \backslash \beta})^{3 / 2} \nu^{\alpha \backslash \beta}_{0}\\
&\nu^{\alpha \backslash \beta}_{\perp} \approx (8/3 \sqrt{\pi})(x^{\alpha \backslash \beta})^{1/2} \nu^{\alpha \backslash \beta}_{0}\\
&\nu^{\alpha \backslash \beta}_{\|} \approx (4/3 \sqrt{\pi})(x^{\alpha \backslash \beta})^{1/2} \nu^{\alpha \backslash \beta}_{0}\\
&\nu^{\alpha \backslash \beta}_{\epsilon} \approx -(4/\sqrt{\pi}) (x^{\alpha \backslash \beta})^{1 / 2} \nu^{\alpha \backslash \beta}_{0}\\
  &\mathrm{For}\;x^{\alpha \backslash \beta} \gg 1 : \nonumber \\  
&\nu^{\alpha \backslash \beta}_{s} \approx \left ( 1 + m_{\alpha} / m_{\beta} \right ) \nu^{\alpha \backslash \beta}_{0}\\
&\nu^{\alpha \backslash \beta}_{\perp} \approx 2 \nu^{\alpha \backslash \beta}_{0}\\
&\nu^{\alpha \backslash \beta}_{\|} \approx \left ( 1 / x^{\alpha \backslash \beta} \right ) \nu^{\alpha \backslash \beta}_{0}\\
&\nu^{\alpha \backslash \beta}_{\epsilon} \approx 2 \left ( m_{\alpha} / m_{\beta} \right ) \nu^{\alpha \backslash \beta}_{0}
\end{align}

\begin{figure}[h]
\begin{center}
\includegraphics[width=0.75\textwidth]{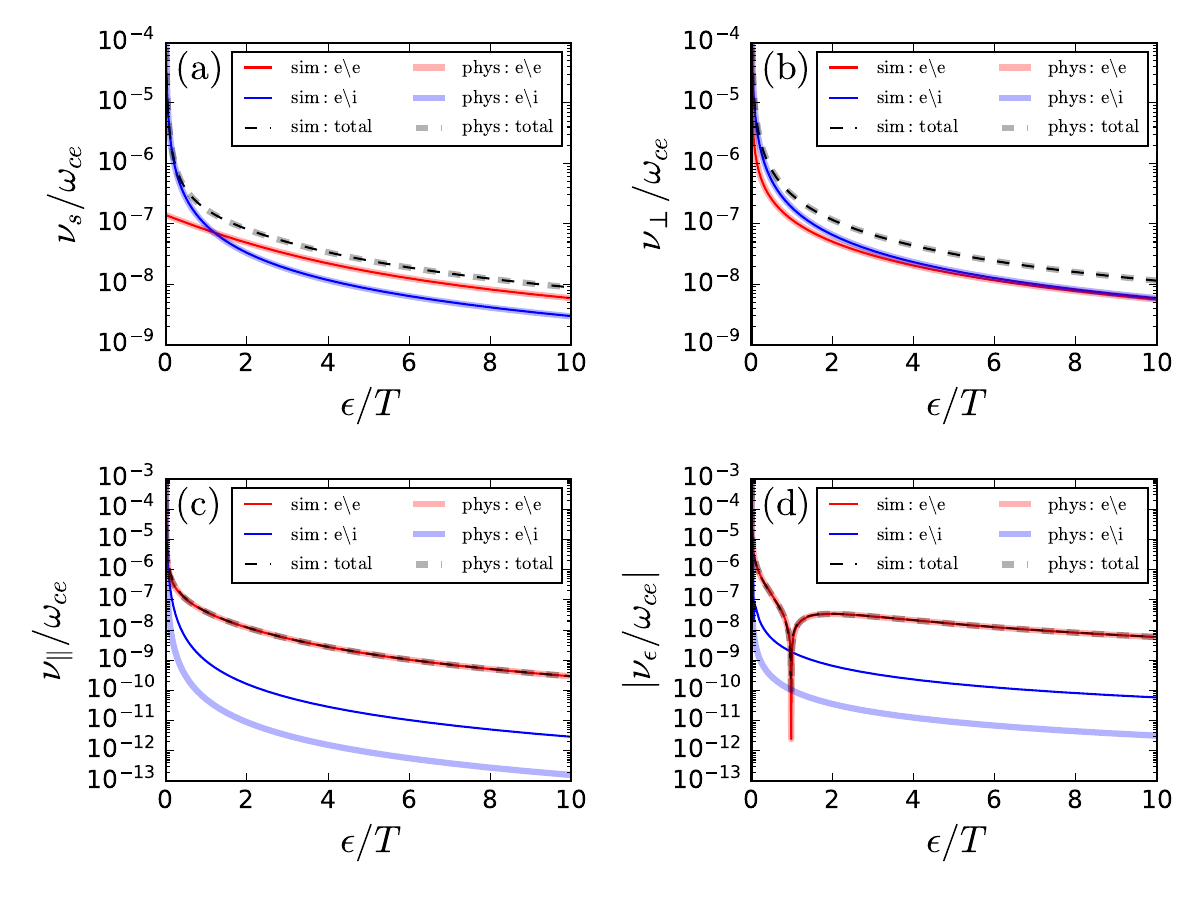}
\caption{\label{fig:scaling_electrons} Electron
relaxation rates as a function of energy. The semi-transparent thick lines show the values of
the physical system, while the thin lines show the values of a system with
an artificial mass ratio that uses our new scaling.
}
\end{center}
\end{figure}

For many laboratory systems, the conditions of
interest are $m_{i}T_{e}/m_{e}T_{i}\gg1$, where the majority of the
collisions will satisfy
$x^{e \backslash i} = m_{i} v_{e}^{2} / 2 k T_{i}\gg 1$ for electron scattering and $x^{i \backslash e} = m_{e} v_{i}^{2} / 2 k T_{e}\ll 1$ for ion scattering. Taking these limits along with $m_{i}/m_{e}\gg1$ and 
examining the scaling of the resulting approximate relaxation rates gives
\begin{align}
  &\mathrm{Electrons} \; (x^{e \backslash i} \gg 1): \nonumber \\
&(\nu^{e \backslash i}_{s})_{sim} / (\nu^{e \backslash i}_{s})_{phys} = 
K_{\nu ei}K_{c}^{4}K_{m}^{-1/2}\\
&(\nu^{e \backslash i}_{\perp})_{sim} / (\nu^{e \backslash i}_{\perp})_{phys} = 
K_{\nu ei}K_{c}^{4}K_{m}^{-1/2}\\
&(\nu^{e \backslash i}_{\|})_{sim} / (\nu^{e \backslash i}_{\|})_{phys} = 
K_{\nu ei}K_{c}^{4}K_{m}^{1/2}\\
&(\nu^{e \backslash i}_{\epsilon})_{sim} / (\nu^{e \backslash i}_{\epsilon})_{phys} = 
K_{\nu ei}K_{c}^{4}K_{m}^{1/2}\\
  &\mathrm{Ions} \;(x^{i \backslash e} \ll 1): \nonumber \\  
&(\nu^{i \backslash e}_{s})_{sim} / (\nu^{i \backslash e}_{s})_{phys} = 
K_{\nu ie}K_{c}^{4}K_{m}^{1/2}\\
&(\nu^{i \backslash e}_{\perp})_{sim} / (\nu^{i \backslash e}_{\perp})_{phys} = 
K_{\nu ie}K_{c}^{4}K_{m}^{1/2}\\
&(\nu^{i \backslash e}_{\|})_{sim} / (\nu^{i \backslash e}_{\|})_{phys} = 
K_{\nu ie}K_{c}^{4}K_{m}^{1/2}\\
&(\nu^{i \backslash e}_{\epsilon})_{sim} / (\nu^{i \backslash e}_{\epsilon})_{phys} = 
K_{\nu ie}K_{c}^{4}K_{m}^{1/2}
\end{align}
To satisfy energy and momentum conservation,
the relaxation rates must be scaled equivalently as
$K_{\nu e i} = K_{\nu i e}$.
Choosing $K_{\nu ei}=K_{c}^{-4}K_{m}^{-1/2}$ will preserve all of the
relaxation rates for the ions.  For the electrons, $\nu^{e \backslash i}_{s}$
and $\nu^{e \backslash i}_{\perp}$ will be matched to the electron
electromagnetic timescales, consistent with the electron-electron contribution
to those rates. $\nu^{e \backslash i}_{\epsilon}$ and $\nu^{e \backslash i}_{\|}$ will instead be matched to the ion time scale (i.e. the characteristic
timescale of the physical system), which has the important result that the
timescale of energy exchange between electrons and ions will be
matched to the physical system.  For these two rates the electron-electron
contribution is typically dominant, so their total values ($\nu^{e}_{\|}=\nu^{e \backslash e}_{\|}+\nu^{e \backslash i}_{\|}$ and $\nu^{e}_{\epsilon}=\nu^{e \backslash e}_{\epsilon}+\nu^{e \backslash i}_{\epsilon}$) are still matched to the
electron electromagnetic timescales (Figure \ref{fig:scaling_electrons}).

\begin{figure}[h]
\begin{center}
\includegraphics[width=0.75\textwidth]{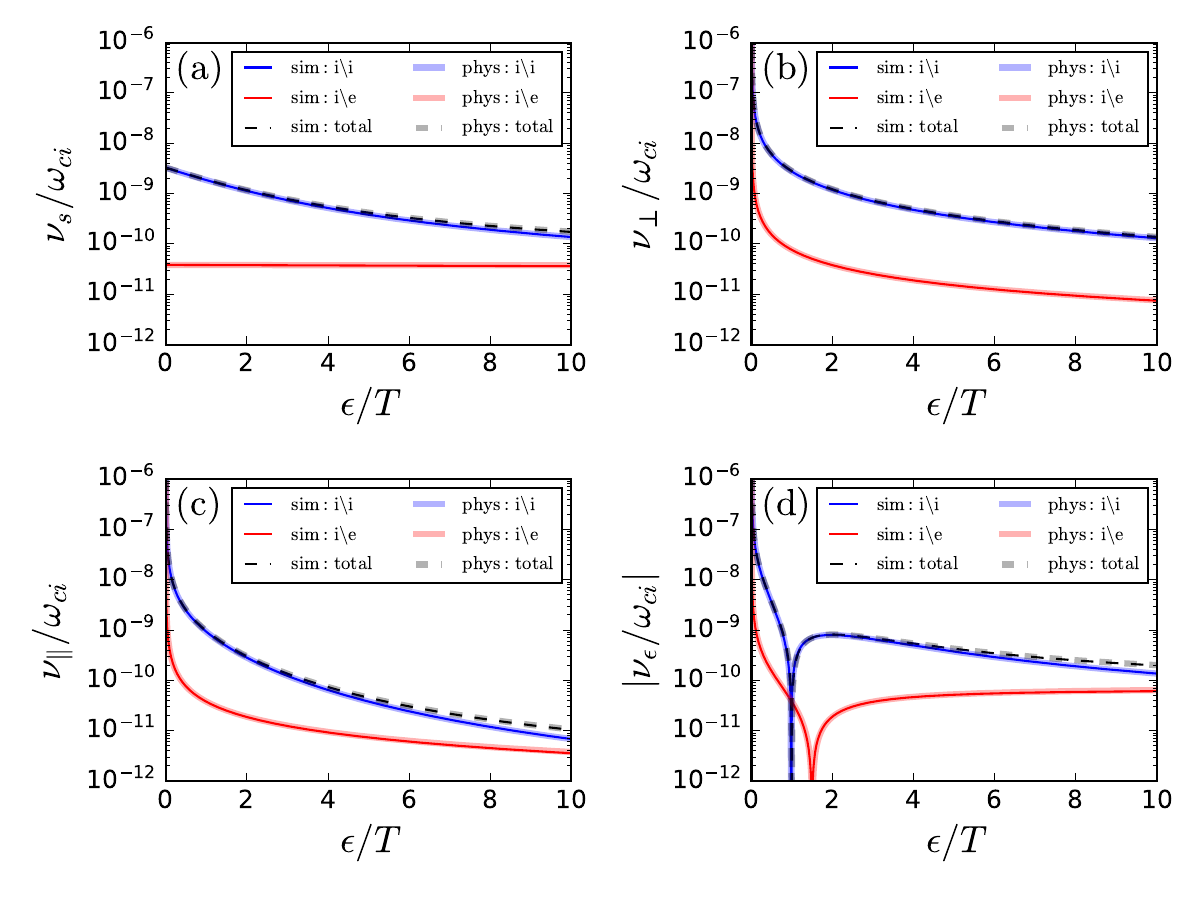}
\caption{\label{fig:scaling_ions} Ion 
relaxation rates as a function of energy. The semi-transparent thick lines show the values of
the physical system, while the thin lines show the values of a system with
an artificial mass ratio that uses our new scaling.
}
\end{center}
\end{figure}

Figures \ref{fig:scaling_ions} and \ref{fig:scaling_electrons} shows the 
results of this scaling for an example electron-proton plasma with 
$T_{e}=T_{i}$ and $(m_{i}/m_{e})_{sim}=100$, 
where the thin lines and thick semi-transparent lines show relaxation 
rates in the simulated
and physical systems, respectively, and the rates for each species are
normalized to their respective cyclotron frequency.
The scaling obtained from the above limiting forms of the expressions are
seen to accurately reproduce all of the total relaxation rates over a wide
energy range. The only significant discrepancies are seen for
$\nu^{e \backslash i}_{||}$ and $\nu^{e \backslash i}_{\epsilon}$, which
as noted above, is due to the electron-ion energy exchange being captured
on the ion timescale (a desirable property for comparison with the physical system),
and does not impact the total value of these rates.

Examining the scaling of the relaxation rates in the opposite
limits ($x^{e \backslash i} \ll 1$ and $x^{i \backslash e} \gg 1$) gives:
\begin{align}
  &\mathrm{Electrons} \; (x^{e \backslash i} \ll 1):\\
&(\nu^{e \backslash i}_{s})_{sim} / (\nu^{e \backslash i}_{s})_{phys} = 
K_{\nu ei}K_{c}^{4}K_{m}^{-2}\\
&(\nu^{e \backslash i}_{\perp})_{sim} / (\nu^{e \backslash i}_{\perp})_{phys} = 
K_{\nu ei}K_{c}^{4}K_{m}^{-1}\\
&(\nu^{e \backslash i}_{\|})_{sim} / (\nu^{e \backslash i}_{\|})_{phys} = 
K_{\nu ei}K_{c}^{4}K_{m}^{-1}\\
&(\nu^{e \backslash i}_{\epsilon})_{sim} / (\nu^{e \backslash i}_{\epsilon})_{phys} = 
K_{\nu ei}K_{c}^{4}K_{m}^{-1}\\
  &\mathrm{Ions} \;(x^{i \backslash e} \gg 1):\\  
&(\nu^{i \backslash e}_{s})_{sim} / (\nu^{i \backslash e}_{s})_{phys} = 
K_{\nu ei}K_{c}^{4}K_{m}^{-1}\\
&(\nu^{i \backslash e}_{\perp})_{sim} / (\nu^{i \backslash e}_{\perp})_{phys} = 
K_{\nu ei}K_{c}^{4}\\
&(\nu^{i \backslash e}_{\|})_{sim} / (\nu^{i \backslash e}_{\|})_{phys} = 
K_{\nu ei}K_{c}^{4}K_{m}^{-1}\\
&(\nu^{i \backslash e}_{\epsilon})_{sim} / (\nu^{i \backslash e}_{\epsilon})_{phys} = 
K_{\nu ei}K_{c}^{4}K_{m}^{-1}
\end{align}

In the low energy limit for electrons we see the scaling
$K_{\nu ei}=K_{c}^{-4}K_{m}^{-1/2}$ will underestimate
the collisionality for all rates compared to both the electron
and ion electromagnetic time scales. However, electrons with energies on the order of $(m_{e} / m_{i})T_{i}$ and below will 
typically be a very small fraction of the total 
distribution and can be expected to play a negligible role in 
the overall dynamics, and therefore it is not a high priority
to match the collisional properties in this regime. For ions,
the high energy limit is relevant when the system features
nonthermal ion acceleration, for which the measurement
is a goal of future laboratory astrophysics experiments of
magnetic reconnection and collisionless shocks. We see
that in the high energy limit for ions our choice of scaling 
will still underestimate the collisionality. However, it
can be expected to be accurate up to energies on the order
of $(m_{i} / m_{e})T_{e}$, which is likely sufficient for
many systems.  Additionally, as all rates drop with increasing
energy, the collisionality may already be negligible over the
timescale of the simulation at energies above
$(m_{i} / m_{e})T_{e}$. We note that this discrepancy at 
the highest energies is still an improvement over past 
approaches that match the timescale of electron-ion energy
exchange to the physical system and scale all collision 
rates in the same way, which even further underestimates the collisionality at all energies
(Figure \ref{fig:collisions_ions}).

In summary, by separately scaling the collisionality of the system according
to $K_{\nu ii}=K_{c}^{-4}$,
$K_{\nu ei}=K_{c}^{-4}K_{m}^{-1/2}$, and $K_{\nu ee}=K_{c}^{-4}K_{m}^{-1/2}$, the total relaxation rates of electrons and ions
will be matched to their electromagnetic time scales, preserving
the properties of particle orbits, and the energy exchange between
electrons and ions will be matched to the ion time scale. 
For simulations
of electromagnetic plasmas with artificial speed of light and mass ratio, this is arguably the best possible choice for reproducing the overall transport 
properties of the system.
This scaling procedure is straightforwardly generalized
to the case with multiple ion species and an 
artificially heavy electron, with the result that 
collisions involving only ions of like or unlike species are scaled as
$K_{\nu i i^{\prime}} = K_{c}^{-4}$, 
collisions between electrons and any ion species are scaled as
$K_{\nu e i} = K_{c}^{-4}K_{m}^{-1/2}$, and collisions
between electrons are scaled as $K_{\nu e e} = K_{c}^{-4}K_{m}^{-1/2}$.
For implementations with a spatially varying Coulomb logarithm that
depends on the local plasma conditions, the values
predicted for the physical system by the simulation should be used in the 
calculation.

\section{Connecting to fluid theory}

The transport equations for a magnetized, collisional multi-component plasma are given by \cite{Braginskii1965,Huba2013NRLFORMULARY}

\begin{equation}\frac{d^{\alpha} n_{\alpha}}{dt} + n_{\alpha} \nabla \cdot {\bf v}_{\alpha} = 0\end{equation}

\begin{equation}m_{\alpha}n_{\alpha}\frac{d^{\alpha} {\bf v}_{\alpha}}{dt} = - \nabla p_{\alpha} - \nabla \cdot P_{\alpha} + Z_{\alpha} e n_{\alpha} \left ( {\bf E} + \frac{1}{c}{\bf v}_{\alpha}\times{\bf B} \right ) + {\bf R}_{\alpha}\end{equation}

\begin{equation}\frac{3}{2} n_{\alpha} \frac{d^{\alpha} T_{\alpha}}{dt} + p_{\alpha} \nabla \cdot {\bf v}_{\alpha} = - \nabla \cdot {\bf q}_{\alpha} - P_{\alpha} : \nabla {\bf v}_{\alpha} + Q_{\alpha}\end{equation}
governing continuity, momentum transport, and energy 
transport, respectively, for the species $\alpha$. The full
definitions of the terms on the right hand sides of the 
momentum and energy transport equations are given in 
\cite{Braginskii1965,Huba2013NRLFORMULARY}. For an electron-ion plasma, the transport 
coefficients depend on the characteristic collision times
$\tau_{e}$ for electrons and $\tau_{i}$ for ions:
\begin{equation}\tau_{e} = \frac{3 \sqrt{m_{e}}T_{e}^{3/2}}{4 \sqrt{2 \pi} \lambda_{ei} e^{4} Z^{2} n_{i}}\end{equation}
\begin{equation}\tau_{i} = \frac{3 \sqrt{m_{i}}T_{i}^{3/2}}{4 \sqrt{\pi} \lambda_{ii} e^{4} Z^{4} n_{i}}\end{equation}
with $\lambda_{ee} = \lambda_{ei}$ assumed.  With the exception of electron viscosity and the left hand side of the electron momentum equation, the equations only
depend on $\tau_{e}$ and $m_{e}$ in the form $\tau_{e} / m_{e}$.
As an example, for an electron-ion plasma with $Z=1$ the momentum transfer to
the electrons from the ions is given by
\begin{equation}{\bf R}_{e} = {\bf R}_{ei} = {\bf R}_{\bf u} + {\bf R}_{T}
\end{equation}
\begin{equation}{\bf R}_{\bf u} = (\tau_{e}/m_{e})^{-1}({\bf j}_{\|}/e)/1.96 + (\tau_{e}/m_{e})({\bf j}_{\perp})(eB^{2}/c^{2})
\end{equation}
\begin{equation}{\bf R}_{T} = -0.71 n \nabla_{\|}T_{e} - (\tau_{e} / m_{e})^{-1}\left ( \frac{3 n c}{2 e B^{2}}\right ) ({\bf B}\times \nabla_{\perp}T_{e})
\end{equation}
where ${\bf R}_{\bf u}$ is the frictional force, ${\bf R}_{T}$ is the
thermal force, and ${\bf j}_{\|}$ and ${\bf j}_{\perp}$ are the electrical
currents parallel and perpendicular, respectively, to the magnetic field 
${\bf B}$.

The scaling of the collisional terms with artificial mass ratio
and reduced speed of light are then
\begin{equation}(\tau_{e} / m_{e})_{sim}/(\tau_{e} / m_{e})_{phys} = (K_{\nu ei} K_{c}^{4} K_{m}^{1/2})^{-1}\end{equation}
\begin{equation}(\tau_{i})_{sim}/(\tau_{i})_{phys} = (K_{\nu ii} K_{c}^{4})^{-1}\end{equation}
We see that our suggested choices of $K_{\nu ei} = K_{\nu ee} = K_{c}^{-4}K_{m}^{-1/2}$ and $K_{\nu ii} = K_{c}^{-4}$ leave
these quantities invariant to the scaling.
The remaining dependencies on 
$e$ and $c$ are all invariant to the speed of light scaling, with the
possible exception of the electric field term in the force equation.
The scaling of the electric field term is
\begin{equation}(Z_{\alpha} e n_{\alpha} {\bf E})_{sim} / (Z_{\alpha} e n_{\alpha} {\bf E})_{phys} = ({\bf E})_{sim}/({\bf E})_{phys}K_{c}\end{equation}
where we must separately consider the impact of the scaling on
different possible electric field sources. Rearranging the momentum
equation in the form of an Ohm's law gives
\begin{equation}{\bf E} = - \frac{1}{c}{\bf v}_{\alpha}\times{\bf B} + \frac{1}{Z_{\alpha} e n_{\alpha}}\left( \nabla p_{\alpha} + \nabla \cdot P_{\alpha} - {\bf R}_{\alpha} + m_{\alpha}n_{\alpha}\frac{d^{\alpha} {\bf v}_{\alpha}}{dt}\right )\end{equation}
and allows us to examine the scaling of different possible electric
field sources. With the exception of the electron inertial term and the viscosity term,
all of the terms on the right produce electric fields that scale as $({\bf E})_{sim}/({\bf E})_{phys} = K_{c}^{-1}$, which leave the electric field term in
the momentum equation invariant. Therefore, the force on the fluid
from the electric field is expected to be matched for large scale transport processes, while
on the electron electrostatic scales it will be modified by the
heavy electron mass and reduced speed of light.

For the heavy electron mass, the only terms that differ between
the physical system and the scaled system are the left hand 
side of the electron momentum equation and the electron
viscosity terms on the right hand sides of the electron momentum 
equation and energy
equation, which are all increased by a factor of $K_{m}$ in
the scaled system.
The dependence of the left hand side of the electron 
momentum equation on the electron mass has the result
that for a given right hand side of the equation,
the momentum changes are matched, \begin{equation}\left ( \frac{d (m_{e} {\bf v}_{e})}{dt} \right )_{sim} = \left ( \frac{d (m_{e} {\bf v}_{e})}{dt} \right )_{phys}\end{equation} while the velocity
changes occur slower by a factor of $K_{m}$ in the
simulated system, \begin{equation}\left ( \frac{d {\bf v}_{e}}{d (\omega_{ce}t)} \right )_{sim} = \left ( \frac{d {\bf v}_{e}}{d(\omega_{ce}t)} \right )_{phys}\end{equation} i.e., they
are matched on the electron electromagnetic timescale.
The electron viscosity coefficients are all increased
by a factor $K_{m}$ in the simulated system with this scaling.  The 
viscosity
includes collisionless and collisional terms, and thus it is not
possible to simultaneously match all of the viscosity coefficients
to the physical system for any choice of scaled collisionality. 
However, the scaling of $K_{\nu ei} = K_{\nu ee} = K_{c}^{-4}K_{m}^{-1/2}$ is the only one that will
preserve the ratio between the collisionless and collisional terms.
The ion coefficients dominate the viscosity for the typical scenario 
of $T_{i} < (m_{i} / m_{e})^{1/2} T_{e}$, so the large scale turbulence in the
simulated system should not be affected by the relatively small increase in
electron viscosity resulting from increased electron mass. 

In the further reduced model of magnetohydrodynamics, important dimensionless 
quantities that characterize the properties of the plasma include the Lundquist number,
$S = L V_{A} / \eta$, and the Prandtl number, $Pr = \nu / \eta$.
Here $V_{A}$ is the Alfven speed, $\nu$ is the kinematic viscosity
and $\eta = c^{2} / 4 \pi \sigma_{0}$ is the magnetic diffusivity, where
$\sigma_{0} = f(Z) n e^{2} \tau_{e} / m_{e}$ is the electrical 
conductivity.  In the scaled system, $L$ is unchanged and 
$V_{A} = B / \sqrt{4 \pi m_{i} n_{i}}$ is invariant.  The magnetic
diffusivity scales as
\begin{equation}\eta_{sim}/\eta_{phys}=(c^{2} m_{e} / e^{2} \tau_{e})_{sim}/(c^{2} m_{e} / e^{2} \tau_{e})_{phys}\end{equation}
which is invariant for the prescribed
scaling (as well as for past approaches that did not consider ion-ion 
collisions \cite{Daughton2009,Daughton2009a}).  As discussed above, the viscosity
$\nu$ is usually dominated by ions and thus not significantly changed
with the mass ratio.  Therefore, in the scaled system both the
Lundquist number and the Prandtl number are approximately invariant to the 
artificial speed of light and mass ratio using our proposed scaling.

\section{Generation of runaway electrons}

For characterizing the physics of runaway electrons, there are
three electric field strengths of interest.
The Dreicer electric field is the strength above which the
runaway of the bulk electron distribution will occur.  This
field is given by $E_{D} = e \lambda_{ei} / \lambda_{D}^{2}$ 
where $\lambda_{D}$ is the Debye length \cite{Dreicer1959ElectronI}.
For nonrelativistic energies, a limited portion of the electron 
distribution will still experience runaway for $E<E_{D}$, and the 
critical electric field allowing runaway of particles with energy 
$\epsilon$ travelling in the direction of the electric field is
$E_{c} = 3 T_{e} E_{D} / \epsilon$ \cite{Dreicer1960ElectronII}.
When relativistic effects are taken into account, it is found
that electric fields with strength greater than
$E_{R} = (T_{e} / m_{e}c^{2}) E_{D}$ cause runaway of a limited
portion of the electron distribution \cite{Connor1975RELATIVISTICELECTRONS}. % Add the expression
To take into account the reduction in $e$, we examine the scaling of the
electric field force as $e E$.
The dependencies of these forces on the artificial physical parameters are given by:

\begin{equation}(e E_{D})_{sim} / (e E_{D})_{phys} = K_{\nu ei}K_{c}^{4}\end{equation}

\begin{equation}(e E_{c})_{sim} / (e E_{c})_{phys} = K_{\nu ei}K_{c}^{4}\end{equation}

\begin{equation}(e E_{R})_{sim} / (e E_{R})_{phys} = K_{\nu ei}K_{c}^{2}K_{m}^{-1}\end{equation}

and thus using the scaling $K_{\nu ei} = K_{m}^{-1/2}K_{c}^{-4}$ we find
\begin{equation}(e E_{D})_{sim} / (e E_{D})_{phys} = K_{m}^{-1/2}\end{equation}

\begin{equation}(e E_{c})_{sim} / (e E_{c})_{phys} = K_{m}^{-1/2}\end{equation}

\begin{equation}(e E_{R})_{sim} / (e E_{R})_{phys} = K_{c}^{-2} K_{m}^{-3/2}\end{equation}

The artificial lowering of the Dreicer field may have significant implications
for the modeling of reconnection with artificial mass ratio \cite{Daughton2009,Daughton2009a}.  This may be
expected to allow for a transition to fast reconnection and nonthermal
particle acceleration at a lower critical Lundquist number in the
simulation system compared to the critical Lundquist number of the physical
system.
We see that this choice of scaling will artificially reduce both the Dreicer
bulk and tail runaway critical fields by a factor $K_{m}^{-1/2}$.
The change in the relativistic tail runaway field $K_{m}^{-3/2}K_{c}^{-2}$
may be less than or greater than one in the range of reasonable
$K_{m}$ and $K_{c}$, and should be checked for a given simulation
to ensure it is not allowing for artificial electron 
acceleration. We conclude that the physics of runaway 
electrons is in general only matched at the full mass 
ratio. Nevertheless, it may be possible to obtain 
important insights and even converged results with 
reduced mass ratio, although this will be
problem-dependent.

\section{Limitations on values of artificial parameters} \label{sec:limitations}

The theory of cumulative small angle scattering is valid when 
collision frequencies satisfy $\nu_{c} / \omega_{p} \ll 1$. For the
typical case where electrons provide the strongest constraint,
the scaled version of the condition on collisionality is given by
$(\nu_{0}^{e \backslash e} / \omega_{pe})_{sim} / (\nu_{0}^{e \backslash e} / \omega_{pe})_{phys} = K_{\nu ee} K_{c}^{3}$.
For the prescribed scaling where the electron collisionality is matched
to the electron cyclotron frequency, we must choose
$K_{\nu ee} = K_{m}^{-1/2} K_{c}^{-4}$.  This will change
the ratio of the collision frequency to the plasma frequency in the
simulation by a factor $K_{m}^{-1/2} K_{c}^{-1}$, which will limit the possible values of $K_{m}$ and 
$K_{c}$ that can be used in a kinetic simulation of a given physical
system as $(\nu_{0}^{e \backslash e} / \omega_{pe})_{phys}  K_{m}^{-1/2} K_{c}^{-1} \ll 1$.  Additionally, the previously mentioned condition 
$(T_{e} / mc^{2})_{phys}K_{m}^{-1}K_{c}^{-2} \ll 1$ is 
necessary to prevent artificial relativistic effects when 
simulating physical systems with nonrelativistic
temperatures.

\section{Implementation}

Existing PIC codes that incorporate particle collisions may be 
straightforwardly modified to incorporate the new scaling method.
We have implemented the scaling method in the PIC code PSC
\cite{Germaschewski2016,Fox2017,Matteucci2018,Pongkitiwanichakul2021IonInteraction,Lezhnin2021KineticPlasmas,Schaeffer2020KineticPlasmas,Fox2018KineticPlasmas}, which includes particle collisions through a module based on the
Takizuka-Abe algorithm \cite{Takizuka1977}. The collisionality in PSC is controlled by a 
user-set parameter $(\nu dt)_{0}$, which is the characteristic
collision frequency Eq. (\ref{eq:nu0}) for electron-electron collisions
assuming a velocity of $v_{e} = c$, multiplied by the simulation timestep
$dt$ to give a cumulative collisionality.
\begin{equation}
    (\nu dt)_{0} = \nu^{e \backslash e}_{0} \left ( v_{e}=c \right ) dt \\
    = \left ( 4 \pi e^{2} e^{2} \lambda_{e e} n_{e} / m_{e}^{2} c^{3} \right ) dt \\
    = \frac{\lambda_{ee}}{4\pi (n_{e} (c/\omega_{pe})^{3})} \left( \omega_{pe} dt \right )
\end{equation}
Each time the collision module is called, the particles in a given
simulation cell are randomly paired and scattered in the center-of-mass frame of each pair, with the the cumulative collisionality $(\nu dt)^{\alpha\backslash\beta}$
for each pair calculated by modifying $(\nu dt)_{0}$ according to
\begin{equation}
    (\nu dt)^{\alpha \backslash \beta} = \left (\frac{Z_{\alpha}^{2}Z_{\beta}^{2} (n / n_{0}) N_{collide}}{m_{\alpha\beta}v_{rel}^{3}} \right ) (\nu dt)_{0}
\end{equation}
where $Z_{\alpha}$ and $Z_{\beta}$ are the charges of species $\alpha$ and $\beta$ normalized to the electron charge $e$, $n/n_{0}$ takes into account
variations in the local density from the characteristic density $n_{0}$, 
$m_{\alpha\beta}=m_{\alpha}m_{\beta}/(m_{\alpha}+m_{\beta})$ is the
reduced mass of the pair, and $v_{rel}$ is the relative velocity of the
pair in the center-of-mass frame.  Additionally, the user-set parameter $N_{collide}$
is included to allow for subcycling of the collision module every $N_{collide}$
timesteps. This subcycling allows for computational savings while retaining
accuracy as long as $(\nu dt)^{\alpha \backslash \beta} \ll 1$.

The momenta of the particles in each collision pair are rotated by a polar
angle $\theta$ and azimuthal angle $\psi$ relative to the incident velocity vectors, utilizing two numbers $r_{1}$
and $r_{2}$ randomly sampled from the unit interval $[0, 1)$.
The azimuthal angle is randomly chosen between 0 and $2\pi$.
\begin{equation}
    \psi = 2 \pi r_{1}
\end{equation}
If $(\nu dt)^{\alpha \backslash \beta} < 1$, the scattering angle is 
calculated as
\begin{equation}
    \theta = 2\arctan{\left(\sqrt{-(1/2)(\nu dt)^{\alpha \backslash \beta}\mathrm{ln}(1-r_{2})}\right)}
    \label{eq:theta}
\end{equation}
If $(\nu dt)^{\alpha \backslash \beta} \geq 1$, the collision is not
resolved and $\theta$ is calculated to ensure uniform random sampling over the surface of a sphere.
\begin{equation}
    \theta = \arccos{\left(1-2r_{2}\right)}
\end{equation}
For a given simulation, $N_{collide}dt$ must be set sufficiently
small so that such unresolved collisions are rare to ensure the 
collisionality is being accurately modeled.

To implement the new scaling, we assume $(\nu dt)_{0}$ has been set
according to the parameters of the physical system, and new
factors
$\Gamma_{\alpha\beta}$ are defined for each possible pair of particle species $\alpha$ 
and $\beta$. The cumulative collisionality used to calculate the scattering
angle in Eq. (\ref{eq:theta}) for each particle
collision is then modified as
\begin{equation}
    (\nu dt)^{\alpha \backslash \beta}_{scaled} = \Gamma_{\alpha\beta}(\nu dt)^{\alpha \backslash \beta}
\end{equation}
To take into account the scaling factors $K_{\nu\alpha\beta}$ discussed above 
and additionally allow for the use of different Coulomb logarithms for each 
pair of species, the new factors $\Gamma_{\alpha\beta}$ are defined as follows.
\begin{equation}
    \Gamma_{\alpha\beta} = K_{\nu\alpha\beta} (\lambda_{\alpha\beta}/\lambda_{ee})
\end{equation}
The additional computation over the standard algorithm is limited to
testing the charges and masses of the particles in each collision
pair to select the appropriate factor $\Gamma_{\alpha\beta}$, and
a single multiplication of the bare collision rate $(\nu dt)^{\alpha \backslash \beta}$ by this factor. This has negligible impact on the
performance of the code, as the computation for each collision is dominated
by transforming to the center-of-mass frame, calculating the scattering 
angles and outgoing particle momenta, and transforming back to the lab 
frame.

\section{Benchmarking Tests}

\begin{figure}[h]
\begin{center}
\includegraphics[width=0.75\textwidth]{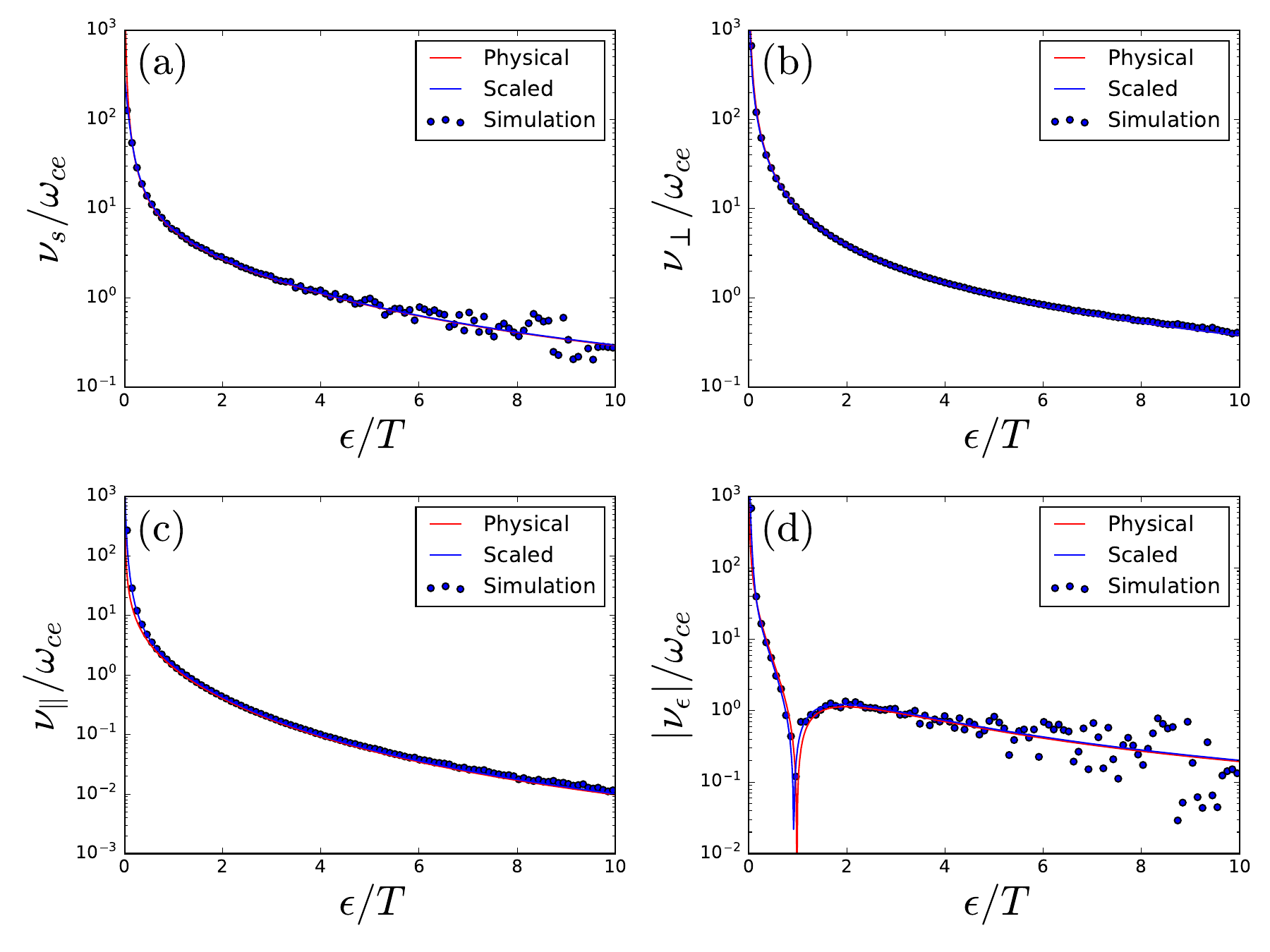}
\caption{\label{fig:collisions_electrons} 
Electron relaxation rates. Solid red line shows rates for the
physical plasma, solid blue line shows rates for the scaled system, and dots show the values measured from the simulation
particles.
}
\end{center}
\end{figure}

\begin{figure}[h]
\begin{center}
\includegraphics[width=0.75\textwidth]{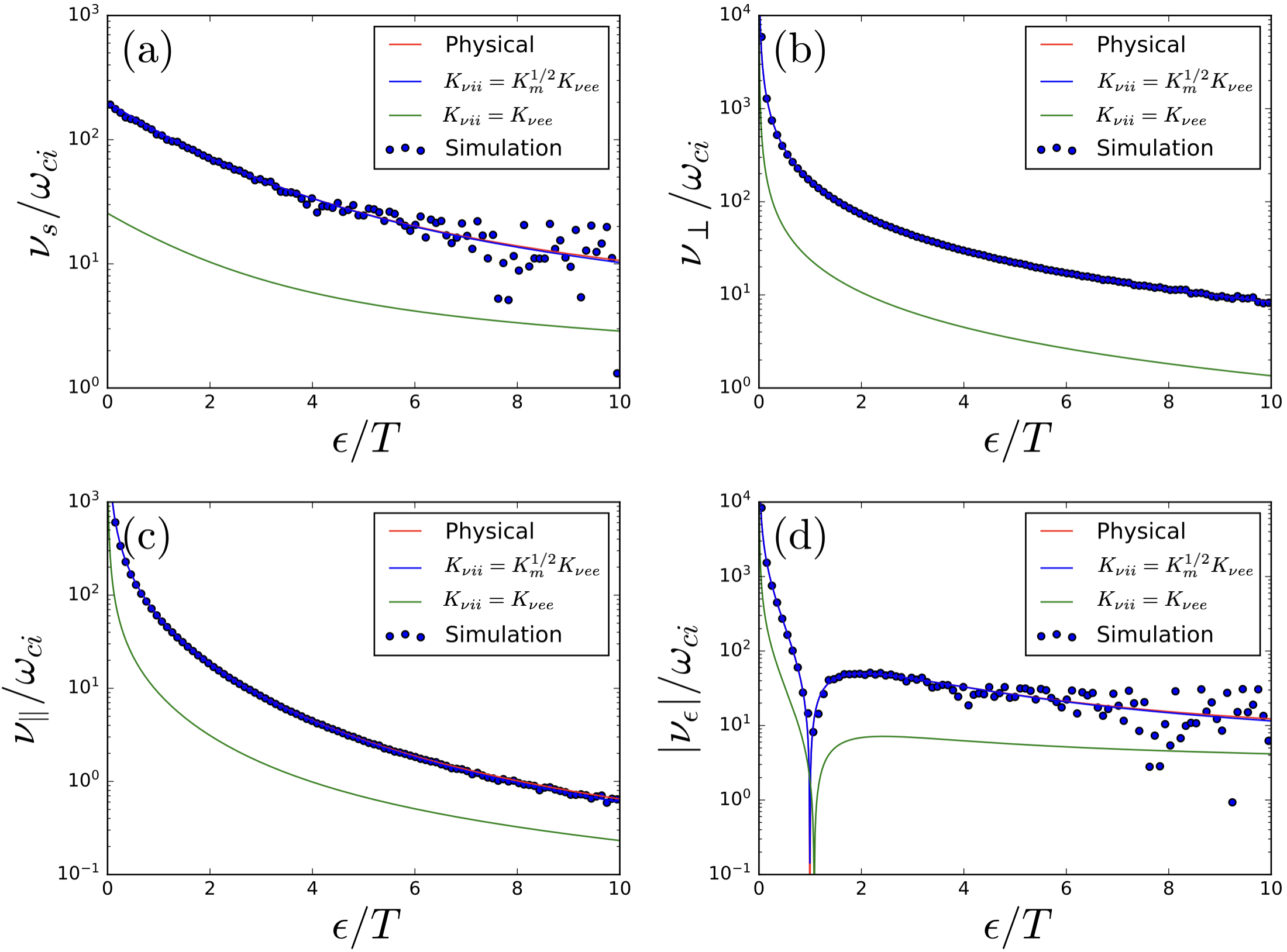}
\caption{\label{fig:collisions_ions} 
Ion relaxation rates. Solid red line shows rates for the
physical plasma, solid blue line shows rates for the scaled system, and dots show the values measured from the simulation
particles. Green line shows the rates that would be obtained
if the electron-ion equilibration time were matched and all
collision rates were scaled in the same way (typical of past 
approaches).
}
\end{center}
\end{figure}

Figures \ref{fig:collisions_electrons} and \ref{fig:collisions_ions} show 
benchmarking tests of the scaled collisionality for
electrons and ions against theoretical test particle relaxation rates
performed using our PSC implementation.
For these tests, we assume the physical system is a charge neutral
electron-proton plasma with electron density $n_{e}= 5 \times
10^{18}\;\mathrm{cm}^{-3}$ and equal electron and ion temperatures of
$T_{e} = T_{i} = 10\;\mathrm{eV}$.  These parameters give a Coulomb
logarithm for electron-electron collisions of $\lambda_{ee} \approx
4.8$, and for simplicity we assume $\lambda_{ee} = \lambda_{ii} =
\lambda_{ei}$.  The simulation is scaled using artificial physical
parameters according to $K_{c} = 1/100$ and $K_{m} = 73.44$, giving
$(m_{i}/m_{e})_{sim} = 25$. The simulation has a time resolution of
$\omega_{pe} dt = 0.0009$, 100 grid cells, 1000 particles-per-cell
per-species, and is evolved for 1000 timesteps.  Both intra- and
inter-species collisions are included in the simulation, and to
isolate the effect of the collision operator the electromagnetic field
solver is turned off.  Average total relaxation rates for each species
are calculated by measuring the change in momentum for each particle
at each timestep, binning by particle energy, and averaging the energy
bins over all timesteps.  Following the discussion above, the
relaxation rates are normalized to the cyclotron frequency for each
species, assuming an arbitrary magnetic field strength of $B =
10\;\mathrm{T}$.  Figure \ref{fig:collisions_electrons} shows the results for electrons, with the
dots showing the values measured by the simulations which show good
agreement with the theoretical values for the scaled system shown by
the solid blue line.  The solid red line shows the theoretical values
corresponding to the physical system, which is nearly
indistinguishable from that for the scaled system. Similar results are
obtained for the ions, plotted in Figure \ref{fig:collisions_ions}. In this plot, the solid
blue line shows the theoretical values for the scaling described above
($K_{\nu ii} = K_{m}^{1/2}K_{\nu ee}$), which agrees closely
with that of the physical system. The additional green curve shows the ion
relaxation rates that would be obtained if one were to apply the same scaling
to all collision rates while ensuring, for example, that the electron-ion
equilibration time is matched to that of the physical system (a common approach), which would significantly
underestimate the ion collisionality.

\begin{figure}[h]
\begin{center}
\includegraphics[width=0.5\textwidth]{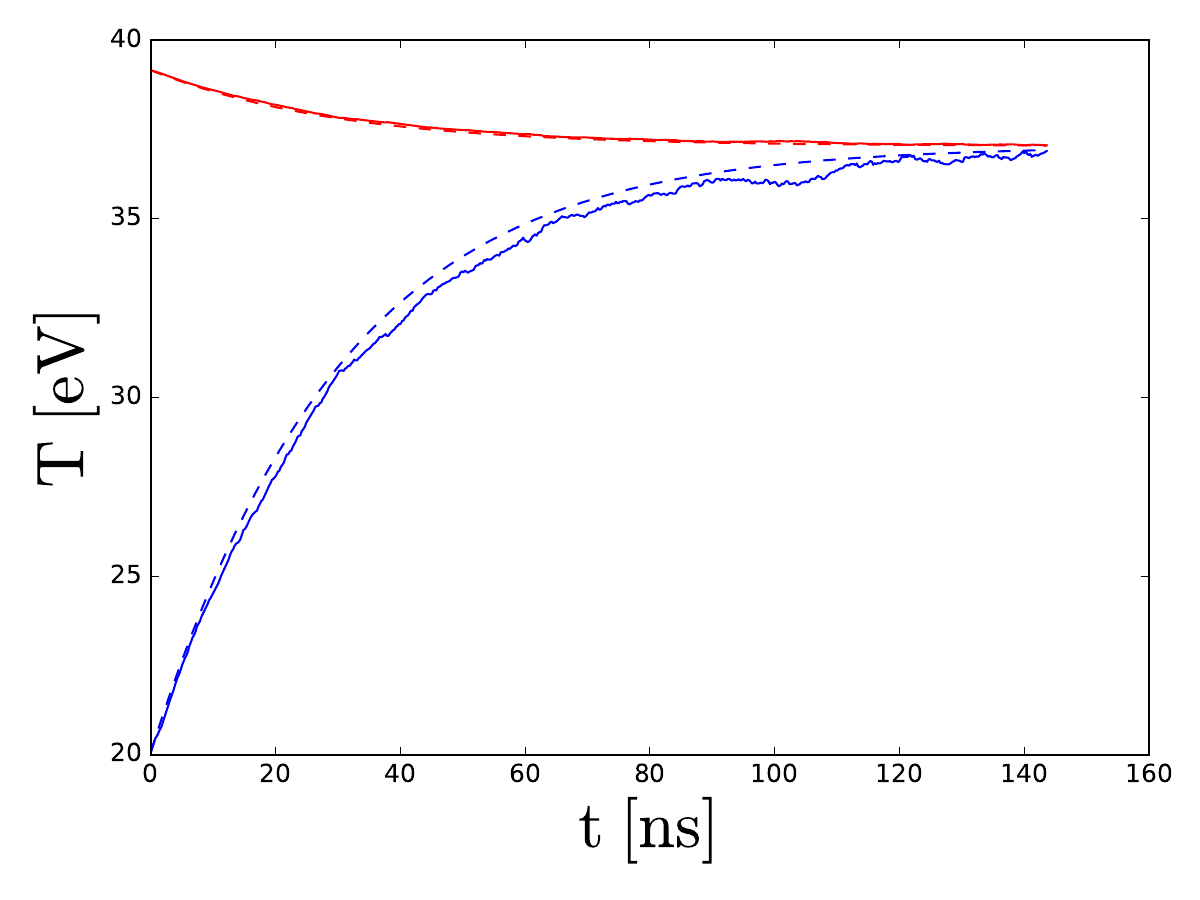}
\caption{\label{fig:equilibration} 
Electron-ion temperature equilibration. Solid red and blue lines show the evolution of the temperatures measured in the
simulation for electrons and ions, respectively, while the
dashed red and blue lines show the theoretical curves for
electrons and ions, respectively.
}
\end{center}
\end{figure}

Figure \ref{fig:equilibration} shows a test of
electron-ion temperature equilibration.  For this
test, the physical plasma has an electron density of
$n_{e}=10^{19}\;\mathrm{cm^{-3}}$, the ions have a charge 
state of $Z=8$ and mass $m_{i}=118 m_{p}$ where $m_{p}$ is the
proton mass, and the initial temperatures are $T_{e} = 40\;\mathrm{eV}$ and $T_{i} = 20\;\mathrm{eV}$. For the simulation, the scaling factors
$K_{c} = 1 / 300$ and $K_{m} = 361.08$ are used.
The solid red and blue lines show the evolution of the
electron and ion temperatures in the simulation, which agree
closely with the theoretical curves shown by the corresponding
dashed red and blue lines. Here the theoretical curves
are calculated as $d T_{e} = \nu_{eq,e} (T_{e}-T_{i}) \Delta t$
and $d T_{i} = \nu_{eq,i} (T_{i}-T_{e}) \Delta t$, where $\nu_{eq,i} = Z \nu_{eq,i}$ and 
\begin{equation}
    \nu_{eq,e} = \frac{Z}{3 \sqrt{2}\pi^{3/2}} \frac{\lambda_{ei}}{(n_{e} d_{e})^{3}}\frac{m_{e}/m_{i}}{(T_{e}+(m_{e}/m_{i})T_{i})^{3/2}}
\end{equation}

\section{Relativistic effects}

While the above analysis was limited to nonrelativistic
collision theory, many natural and laboratory
plasma systems of interest feature relativistic
temperatures and/or relativistic energetic particles.
Multiple models have been developed to extend the 
nonrelativistic algorithm of Takizuka to take into 
account relativistic effects \cite{Sentoku2008,Peano2009,Perez2012}.
However, these algorithms have not yet been benchmarked 
against the full relativistic extensions to the test 
particle relaxation rates and fluid transport 
coefficients, which have only been derived
more recently \cite{Pike2014,Pike2016}. 
It is critical to fully benchmark these collision operators against 
the theoretical formulas to provide relativistic PIC collisions with the 
same firm theoretical grounding as the original nonrelativistic method of 
Takizuka, for which the reproduction of the nonrelativistic particle 
relaxation rates and fluid transport coefficients has been demonstrated.

\begin{figure}[h]
\begin{center}
\includegraphics[width=1.0\textwidth]{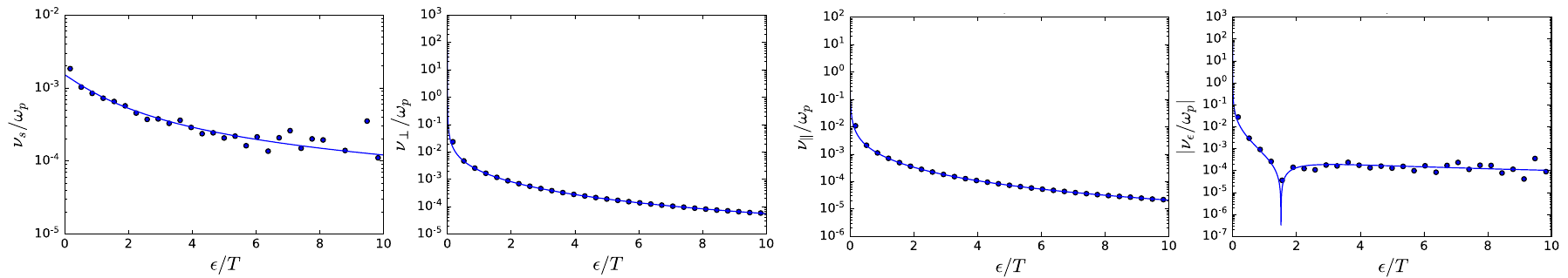}
\caption{\label{fig:collision_rel} 
Benchmarking the relativistic collision algorithm of \cite{Perez2012} against
the full relativistic extensions to the test particle relaxation rates derived
in \cite{Pike2014}. Circles show the energy resolved relaxation rates 
obtained from a PIC simulation of an electron-positron pair plasma with a 
relativistic temperature $T/m_{e}c^{2}=1$, and lines show the predictions of 
the theoretical model.
}
\end{center}
\end{figure}

In preliminary tests for an electron-positron plasma
with a relativistic temperature of $T/m_{e}c^{2}=1$
using the PIC code OSIRIS and the relativistc 
collision model of Perez \cite{Fonseca2013,Perez2012}, we have 
found good agreement between the full relativistic 
extensions to the nonrelativistic test particle relaxation rates (Eq.
(\ref{eq:relaxation_rates})) derived by \cite{Pike2014} 
up to highly relativistic energies
(Figure \ref{fig:collision_rel}). In future work we 
will perform detailed tests of the different
available relativistic collision operators against
the relativistic extensions for the test particle 
relaxation rates and transport coefficients for 
electron-ion plasmas. We will additionally extend 
the scaling arguments presented above into the 
relativistic regime, where the dependence on the 
speed of light now becomes nontrivial.  Even for
simulations that start with 
nonrelativistic temperatures, an artificially slow 
speed of light can enhance relativistic effects for 
electrons, leading to the development of 
artificially relativistic temperatures in small 
regions of the domain or nonthermal particles with
energies that are relativistic only in the scaled 
system. It remains important to 
determine the best approaches for ensuring the 
collisionality of the simulated system represents 
that of the physical system as closely as possible
in the presence of these artificial relativistic effects.

\section{Conclusions}

In conclusion, in this work we have developed a new method
for scaling the collisionality to better preserve the transport
properties in kinetic plasma simulations modeling physical
systems using artificial physical parameters. This scaling
matches the total relaxations rates of electrons and ions
to their respective electromagnetic timescales, and 
additionally preserves important transport properties of the
system including the electron-ion equilibration time. The 
scaling is valid for systems with nonrelativistic temperatures
and $m_{i} T_{e} / m_{e} T_{i} \gg 1$, and will improve the
accuracy of modeling marginally collisional laboratory and
astrophysical systems.  Existing collisional PIC 
implementations can be easily modified to include this
scaling to compensate for artificial physical parameters.
Future work will extend this study to take into account
relativistic effects.

\section*{Acknowledgements}

The research described in this paper was conducted under the Laboratory Directed
Research and Development (LDRD) Program at Princeton Plasma Physics Laboratory,
a national laboratory operated by Princeton University for the U.S. Department 
of Energy under Prime Contract No. DE-AC02-09CH11466.

\bibliographystyle{unsrt}
\bibliography{references}

\begin{thebibliography}{10}

\bibitem{Blandford2014}
Roger Blandford, Paul Simeon, and Yajie Yuan.
\newblock {Cosmic ray origins: An introduction}.
\newblock {\em Nuclear Physics B - Proceedings Supplements}, 256-257:9--22,
  2014.

\bibitem{Birdsall}
C.K. Birdsall and A.B. Langdon.
\newblock {\em {Plasma Physics via Computer Simulation}}.
\newblock McGraw-Hill, 1985.

\bibitem{Takizuka1977}
Tomonori Takizuka and Hirotada Abe.
\newblock {A binary collision model for plasma simulation with a particle
  code}.
\newblock {\em Journal of Computational Physics}, 25(3):205--219, 1977.

\bibitem{Pritchett2001}
P.~L. Pritchett.
\newblock {Geospace Environment Modeling magnetic reconnection challenge:
  Simulations with a full particle electromagnetic code}.
\newblock {\em Journal of Geophysical Research: Space Physics},
  106(A3):3783--3798, 2001.

\bibitem{Hesse2001CollisionlessModeling}
Michael Hesse, Joachim Birn, and Masha Kuznetsova.
\newblock {Collisionless magnetic reconnection: Electron processes and
  transport modeling}.
\newblock {\em Journal of Geophysical {\ldots}}, 106:3721--3735, 2001.

\bibitem{Fox2018KineticPlasmas}
W.~Fox, J.~Matteucci, C.~Moissard, D.~B. Schaeffer, A.~Bhattacharjee,
  K.~Germaschewski, and S.~X. Hu.
\newblock {Kinetic simulation of magnetic field generation and collisionless
  shock formation in expanding laboratory plasmas}.
\newblock {\em Physics of Plasmas}, 25(10), 2018.

\bibitem{Daughton2009}
W.~Daughton, V.~Roytershteyn, B.~J. Albright, H.~Karimabadi, L.~Yin, and
  Kevin~J. Bowers.
\newblock {Influence of Coulomb collisions on the structure of reconnection
  layers}.
\newblock {\em Physics of Plasmas}, 16(7), 2009.

\bibitem{Daughton2009a}
W.~Daughton, V.~Roytershteyn, B.~J. Albright, H.~Karimabadi, L.~Yin, and
  Kevin~J. Bowers.
\newblock {Transition from collisional to kinetic regimes in large-scale
  reconnection layers}.
\newblock {\em Physical Review Letters}, 103(6):1--4, 2009.

\bibitem{Fox2011a}
W.~Fox, A.~Bhattacharjee, and K.~Germaschewski.
\newblock {Fast magnetic reconnection in laser-produced plasma bubbles}.
\newblock {\em Physical Review Letters}, 106(21):1--4, 5 2011.

\bibitem{Fox2012a}
W.~Fox, A.~Bhattacharjee, and K.~Germaschewski.
\newblock {Magnetic reconnection in high-energy-density laser-produced
  plasmas}.
\newblock {\em Physics of Plasmas}, 19(2012):056309, 2012.

\bibitem{Le2015}
A.~Le, J.~Egedal, W.~Daughton, V.~Roytershteyn, H.~Karimabadi, and C.~Forest.
\newblock {Transition in electron physics of magnetic reconnection in weakly
  collisional plasma}.
\newblock {\em Journal of Plasma Physics}, 81(01):305810108, 2015.

\bibitem{Trubnikov1965}
B.~A. Trubnikov.
\newblock {Particle Interactions in a Fully Ionized Plasma}.
\newblock {\em Reviews of Plasma Physics}, 1, 1965.

\bibitem{Huba2013NRLFORMULARY}
J~D Huba.
\newblock {NRL PLASMA FORMULARY}.
\newblock {\em Plasma Physics}, 2013.

\bibitem{Braginskii1965}
S.~I. Braginskii.
\newblock {Transport Processes in a Plasma}.
\newblock {\em Reviews of Plasma Physics}, 1:205, 1965.

\bibitem{Dreicer1959ElectronI}
H.~Dreicer.
\newblock {Electron and ion runaway in a fully ionized gas. I}.
\newblock {\em Physical Review}, 115(2), 1959.

\bibitem{Dreicer1960ElectronII}
H~Dreicer.
\newblock {Electron and. Ion Runaway in a Fully Ionized Gas. II*}.
\newblock Technical report, 1960.

\bibitem{Connor1975RELATIVISTICELECTRONS}
J~W Connor and R~J Hastie.
\newblock {RELATIVISTIC LIMITATIONS ON RUNAWAY ELECTRONS}.
\newblock Technical report, 1975.

\bibitem{Germaschewski2016}
Kai Germaschewski, William Fox, Stephen Abbott, Narges Ahmadi, Kristofor
  Maynard, Liang Wang, Hartmut Ruhl, and Amitava Bhattacharjee.
\newblock {The Plasma Simulation Code: A modern particle-in-cell code with
  patch-based load-balancing}.
\newblock {\em Journal of Computational Physics}, 318:305--326, 2016.

\bibitem{Fox2017}
W.~Fox, J.~Park, W.~Deng, G.~Fiksel, A.~Spitkovsky, and A.~Bhattacharjee.
\newblock {Astrophysical particle acceleration mechanisms in colliding
  magnetized laser-produced plasmas}.
\newblock {\em Physics of Plasmas}, 24(9):092901, 2017.

\bibitem{Matteucci2018}
J.~Matteucci, W.~Fox, A.~Bhattacharjee, D.~B. Schaeffer, C.~Moissard,
  K.~Germaschewski, G.~Fiksel, and S.~X. Hu.
\newblock {Biermann-Battery-Mediated Magnetic Reconnection in 3D Colliding
  Plasmas}.
\newblock {\em Physical Review Letters}, 121(9):1--6, 2018.

\bibitem{Pongkitiwanichakul2021IonInteraction}
Peera Pongkitiwanichakul, William Fox, David Ruffolo, Kittipat Malakit,
  Kirill~V. Lezhnin, Jack Matteucci, and Amitava Bhattacharjee.
\newblock {Ion Acceleration in Driven Magnetic Reconnection during
  High-energy–Density Plasma Interaction}.
\newblock {\em The Astrophysical Journal}, 907(2), 2021.

\bibitem{Lezhnin2021KineticPlasmas}
K.~V. Lezhnin, W.~Fox, D.~B. Schaeffer, A.~Spitkovsky, J.~Matteucci,
  A.~Bhattacharjee, and K.~Germaschewski.
\newblock {Kinetic Simulations of Electron Pre-energization by Magnetized
  Collisionless Shocks in Expanding Laboratory Plasmas}.
\newblock {\em The Astrophysical Journal Letters}, 908(2), 2021.

\bibitem{Schaeffer2020KineticPlasmas}
D.~B. Schaeffer, W.~Fox, J.~Matteucci, K.~V. Lezhnin, A.~Bhattacharjee, and
  K.~Germaschewski.
\newblock {Kinetic simulations of piston-driven collisionless shock formation
  in magnetized laboratory plasmas}.
\newblock {\em Physics of Plasmas}, 27(4):042901, 2020.

\bibitem{Sentoku2008}
Y.~Sentoku and A.~J. Kemp.
\newblock {Numerical methods for particle simulations at extreme densities and
  temperatures: Weighted particles, relativistic collisions and reduced
  currents}.
\newblock {\em Journal of Computational Physics}, 227(14):6846--6861, 2008.

\bibitem{Peano2009}
F.~Peano, M.~Marti, L.~O. Silva, and G.~Coppa.
\newblock {Statistical kinetic treatment of relativistic binary collisions}.
\newblock {\em Physical Review E - Statistical, Nonlinear, and Soft Matter
  Physics}, 79(2):2--5, 2009.

\bibitem{Perez2012}
F.~P{\'{e}}rez, L.~Gremillet, A.~Decoster, M.~Drouin, and E.~Lefebvre.
\newblock {Improved modeling of relativistic collisions and collisional
  ionization in particle-in-cell codes}.
\newblock {\em Physics of Plasmas}, 19(8):083104, 2012.

\bibitem{Pike2014}
O.~J. Pike and S.~J. Rose.
\newblock {Dynamical friction in a relativistic plasma}.
\newblock {\em Physical Review E - Statistical, Nonlinear, and Soft Matter
  Physics}, 89(5), 2014.

\bibitem{Pike2016}
O.~J. Pike and S.~J. Rose.
\newblock {Transport coefficients of a relativistic plasma}.
\newblock {\em Physical Review E}, 93(5), 2016.

\bibitem{Fonseca2013}
R~A Fonseca, J~Vieira, F~Fiuza, A~Davidson, F~S Tsung, W~B Mori, and L~O Silva.
\newblock {Exploiting multi-scale parallelism for large scale numerical
  modelling of laser wakefield accelerators}.
\newblock {\em Plasma Physics and Controlled Fusion}, 55(12):124011, 12 2013.

\end{thebibliography}

\end{document}